\documentclass[twocolumn,showpacs,preprintnumbers,amsmath,amssymb]{revtex4}

\usepackage{graphicx}
\usepackage{dcolumn}
\usepackage{bm}
\usepackage{amsfonts}

\newcommand{\ket}[1]{\mbox{$\mid \! #1 \, \rangle$}}
\newcommand{\bra}[1]{\mbox{$\langle \, #1 \! \mid$}}

\newcommand{\dg}[1]{#1 \,^{\circ}}
\newcommand{\eea}{\end{eqnarray}}
\newcommand{\bea}{\begin{eqnarray}}
\newcommand{\eeas}{\end{eqnarray*}}
\newcommand{\beas}{\begin{eqnarray*}}
\newcommand{\WW}{\ensuremath{\mathcal{W}}}
\newcommand{\CC}{\ensuremath{\mathcal{C}}}

\newcommand{\sx}{\sigma_x}
\newcommand{\sy}{\sigma_y}
\newcommand{\sz}{\sigma_z}
\newcommand{\SZ}{\ensuremath{\mathcal{S}}}

\begin{document}


\title{Experimental Analysis of a 4-Qubit Cluster State}

\author{Nikolai Kiesel$^{1,2}$, Christian Schmid$^{1,2}$, Ulrich Weber$^{1,2}$,
Otfried G\"uhne$^{3}$, G\'eza T\'oth$^{2}$, Rupert Ursin$^{4}$,
Harald Weinfurter$^{1,2}$}

\affiliation{
$^{1}$Max-Planck-Institut f{\"u}r Quantenoptik, D-85748 Garching, Germany\\
$^{2}$Department f\"ur Physik, Ludwig-Maximilians-Universit{\"a}t,
D-80797
M{\"u}nchen, Germany\\
 $^{3}$Institut f\"ur Quantenoptik und Quanteninformation,
\"Osterreichische Akademie der Wissenschaften,  A-6020 Innsbruck,
Austria\\
 $^{4}$Institut f{\"u}r Experimentalphysik,
Universit{\"a}t Wien, A-1090 Wien, Austria}%

\date{\today}

\begin{abstract}

Linear optics quantum logic operations enabled the observation of
a four-photon cluster state. We prove genuine four-partite
entanglement and study its persistency, demonstrating remarkable
differences to the usual GHZ state. Efficient analysis tools are
introduced in the experiment, which will be of great importance in
further studies on multi-particle entangled states.
\end{abstract}


\maketitle

Multipartite entangled states play a fundamental role in the field
of quantum information theory and its applications. Recently,
special types of entangled multiqubit states, the so-called graph
states, have moved into the center of interest \cite{graphstates}.
Due to the fact that they can be generated by next-neighbor
interactions, these states occur naturally in solid state systems
or can be easily obtained in experiments on atomic lattices
\cite{BlochCluster}. These graph states are basic elements of
various quantum error correcting codes \cite{errorcorr} and
multi-party quantum communication protocols \cite{multipartyqc}.
Well known members of this family of states are the GHZ and
cluster states. The latter received a lot of attention in the
context of the so-called one-way quantum computer scheme suggested
by Briegel and Raussendorf \cite{oneway_concept}. There, the
cluster state serves as the initial resource of a universal
computation scheme based on single-qubit operations only. Very
recently the principal feasibility of this approach was
experimentally demonstrated for a four-photon cluster state
\cite{vienna}.

In this letter we report the experimental detection of a high
fidelity four-photon cluster state. The inherent stability of the
linear optics phase gate implemented here allowed a detailed
characterization of the states entanglement properties as well as
of its entanglement persistency under loss of qubits.
Introducing stabilizer formalism \cite{stabilizer} for the
experimental analysis we were able to detect genuine four partite
entanglement and to determine the states fidelity with a minimum
number of measurements.


 The four-qubit cluster state can be written in the form
\begin{multline}
\ket{\CC_4}=\frac{1}{2}(\ket{HHHH}_{abcd}+\ket{HHVV}_{abcd}\\
+\ket{VVHH}_{abcd}-\ket{VVVV}_{abcd}),
 \label{eq:Cluster}
\end{multline}
where $\ket{H_i}$ and $\ket{V_i}$ denote linear horizontal (H) and
vertical (V) polarization of a photon
 in the spatial mode $i$ ($i
=a,b,c,d$). If one compares this state with the product of two
Bell states
$\ket{\Phi^+}_{ij}=\frac{1}{\sqrt{2}}(\ket{HH}_{ij}+\ket{VV}_{ij})$,
one observes that the states are equal up to a phase factor of the
last term. This phase can be generated by a controlled phase
(C-Phase) gate acting on input modes $b'$ and $c'$ as defined by
\begin{equation}
\label{eq:PG} \text{C-Phase gate}:
\begin{cases}
\ket{HH}_{b'c'} \rightarrow  \ket{HH}_{bc} \\ 
\ket{HV}_{b'c'} \rightarrow  \ket{HV}_{bc} \\
\ket{VH}_{b'c'} \rightarrow  \ket{VH}_{bc} \\
\ket{VV}_{b'c'} \rightarrow -\ket{VV}_{bc}\,.
\end{cases}
\end{equation}
This scheme directly reflects the generation principle of graph
states: evidently, the state $\ket{\Phi^+}$ can be generated by
next-neighbor interaction, and the four-qubit cluster state is
then obtained by a third interaction between the neighboring
qubits $b'$ and $c'$.

\begin{figure}
\includegraphics[width=8.5cm,clip]{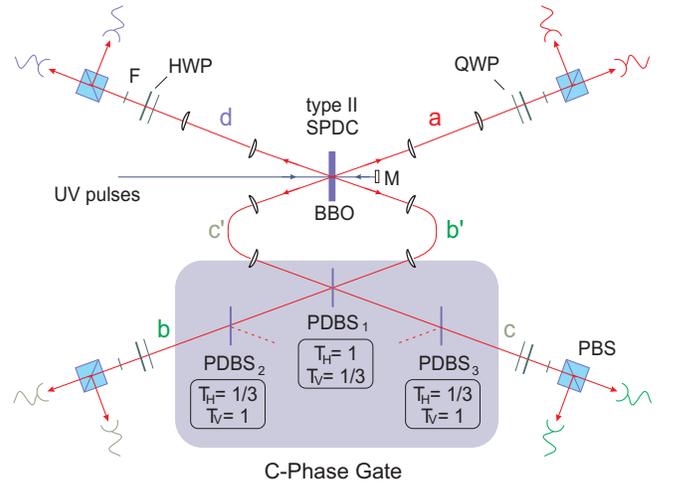}
\caption{Experimental setup for the demonstration of the
four-photon polarization-entangled cluster state. The state is
observed after entangling two EPR-pairs via a linear optics
controlled phase gate (C-Phase gate), which employs two-photon
interference at polarization dependent beam splitters (PDBS). Two
entangled photon pairs are originating from type II spontaneous
parametric down conversion (SPDC) by pumping a $\beta$-Barium
Borate (BBO) crystal in a double pass configuration.  Half- and
quarter wave plates (HWP, QWP) together with polarizing beam
splitters (PBS) are used for the polarization analysis.}
\label{fig:Clustersetup}
\end{figure}

To experimentally implement the C-Phase gate for photons we
simplified the linear-optics gate introduced
recently~\cite{Ralph_PG}, thereby enabling its application in a
four-photon experiment. Since stability is indispensable in
multi-photon experiments we replace (phase-dependent)
single-photon interferometers for different polarizations by a
polarization dependent (but phase-independent) two-photon
interference \cite{Kiesel}. It is well known, that in two-photon
interference both photons leave the same output of the beam
splitter \cite{HOM}. However, if reflectivity and transmittance
are not equal, first, there is a certain probability for the
photons to be detected in different outputs, and second, this term
of the wavefunction might acquire also a phase shift of $\pi$.
Using  a beam splitter with polarization-dependent splitting ratio
(PDBS), one can tune parameters such, that the desired action is
achieved for photons leaving the beam splitter in different output
ports. Optimal action is achieved, if the gate-input photons are
overlapped on a PDBS$_1$, with transmission for horizontal
polarization $T_H=1$, and for vertical polarization $T_V=1/3$. In
order to equalize the transmittance for all input polarizations,
beam splitters (PDBS$_2$ and PDBS$_3$) with the complementary
transmissions ($T_H=1/3$, $T_V=1$) are placed in each output of
the overlap beam splitter (see Fig. \ref{fig:Clustersetup})
\cite{Kiesel}. All together, the detection probability of
coincidences in the output of the gate, and thus the rate of
operation, is $1/9$ independent of the input state.

In the experiment, we use spontaneous parametric down conversion
for the preparation of the two EPR pairs. UV pulses with a central
wavelength of 390~nm and an average power of 700~mW from a
frequency-doubled mode-locked Ti:sapphire laser (pulse length
130~fs) are used in a double pass configuration to pump a 2~mm
thick BBO ($\beta$-Barium Borate, type-II) crystal to obtain two
polarization entangled photon-pairs in distinguishable outputs.
Effects originating from double pair emission into one or the
other pair of outputs are suppressed as all detections are
conditioned to registering one photon in each of the four output
modes $a,b,c$, and $d$. Coupling the four photons into single mode
fibers already behind the BBO-crystal optimizes collection
efficiency and exactly defines the spatial modes, the spectral
selection is achieved with narrow bandwidth interference filters F
($\Delta\lambda = 2$~nm in the C-Phase Gate and $\Delta\lambda =
3$~nm in modes $a$ and $d$) before detection. For the initial
alignment of the C-Phase gate, first, photons originating from one
SPDC process and, second, triggered Hong-Ou-Mandel interference
between the photons emerging from the two SPDC-processes is used
to set the temporal overlap (see Fig. \ref{fig:Clustersetup}).
This setup is sensitive only to length changes on the order of the
coherence length of the detected photons ($\approx150 \mu m$) and
thus stays stable over several days with a typical four-fold
coincidence count rate of 150 per hour.

Polarization analysis is performed in all of the four outputs. All
eight Si-APD single photon counters are fed into a multi-channel
coincidence unit which allows to simultaneously register any
possible coincidence detection between the inputs.  The rates for
each of the 16 characteristic four-fold coincidences have to be
corrected for the difference in the efficiencies of the detectors.
The errors on all quantities are deduced from propagated
Poissonian counting statistics of the raw detection events and
independently determined efficiencies.


\begin{figure}
\includegraphics[width=8cm]{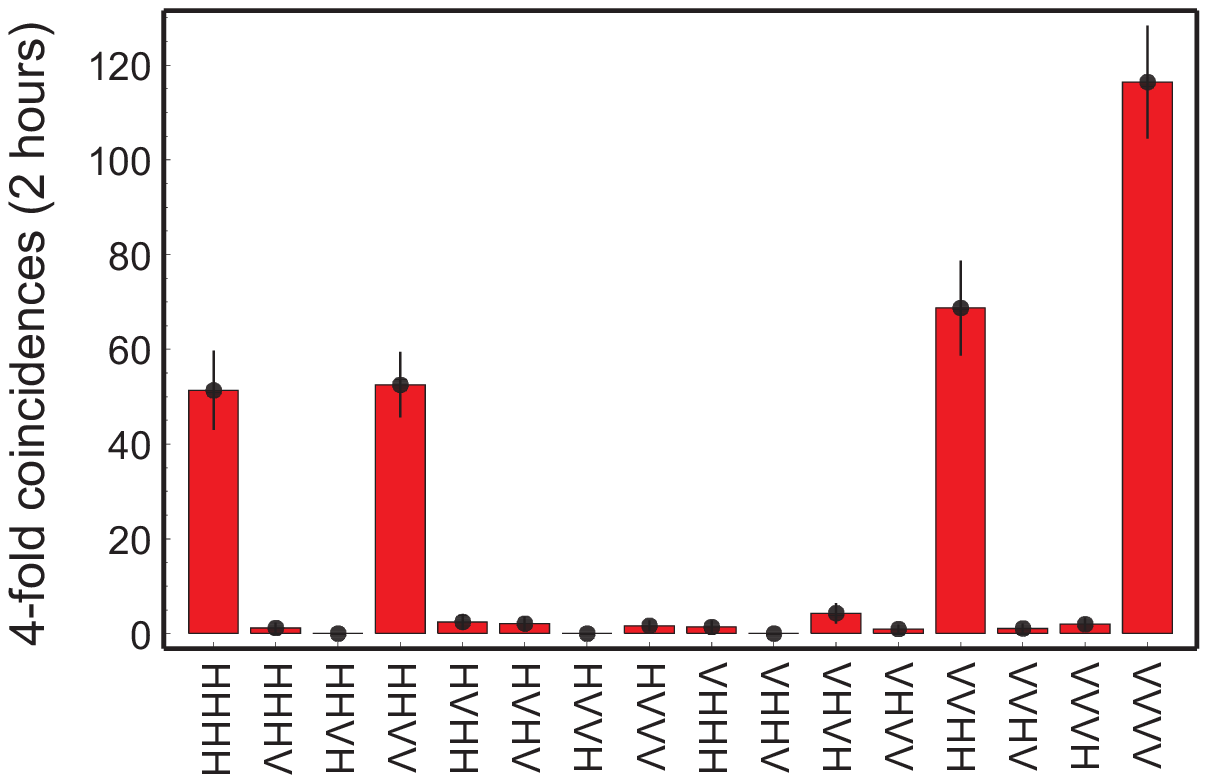}
\caption{Four-fold coincidence counts obtained during two hours of
measurement when analysing in the H/V - basis in all four output
modes.} \label{fig:Clusterstats}
\end{figure}

Figure 2. displays the counts obtained for the four-photon cluster
state (\ref{eq:Cluster}). One clearly observes the four-term
structure with peaks at $HHHH$, $HHVV$, $VVHH$, and $VVVV$. The
$VVVV$-contribution is enhanced, due to non-perfect
indistinguishability of the photons at the overlap beam splitter
in the phase gate, resulting in additional, polarized noise. These
data alone, however, do not prove the contributions to be in a
coherent superposition; various bases have to be analyzed.
Exemplarily we show the four-photon concidence counts when the
photons in mode $a$, $b$ (Fig. 3a), or the photons in mode $c$,
$d$ (Fig. 3b), respectively, are measured along $\dg{\pm45}$. The
clear four term structure is present here as well and indicates
the coherence of the cluster state observed. The imperfect
interference results in an increase of detections with $VVxx$ (a),
or $xxVV$ (b), respectively ($x=+45^\circ/-45^\circ$).

\begin{figure}
\includegraphics[width=8 cm]{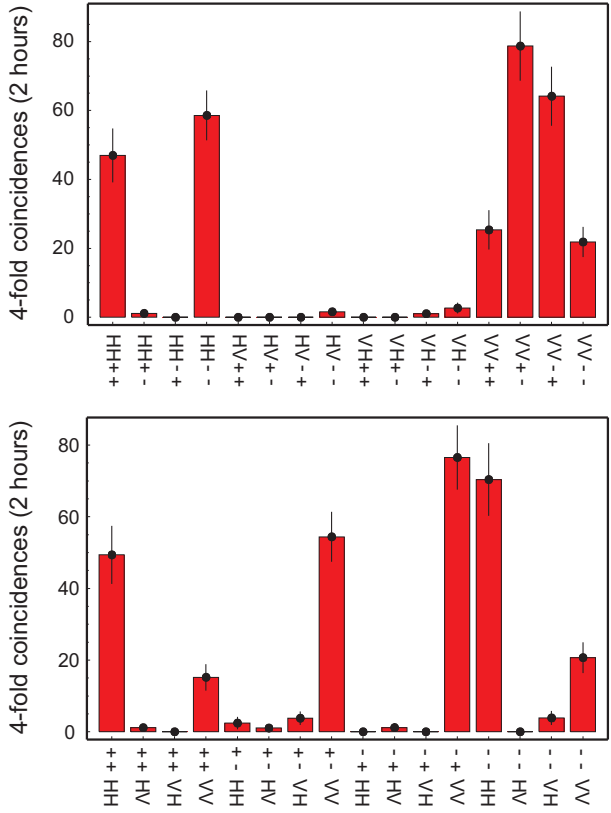}
\caption{Four-fold coincidence counts obtained during two hours of
measurement when analyzing in the H/V-basis in modes $a$ and $b$
and in the $\pm45^\circ$-basis in modes $c$ and $d$ (a), or when
analyzing in the $\pm45^\circ$-basis in modes $a$ and $b$ and in
the H/V-basis in modes $c$ and $d$ (b), respectively. These data
are already sufficient to prove four-photon entanglement based on
stabilizer witnesses.} \label{fig:Clusterstats2}
\end{figure}

Graph states share the common property that an entanglement
witness testing four-partite entanglement \cite{Witness_US} can be
constructed via the so-called stabilizer operators (for the
four-qubit case, see Table I), resulting, for the state
$\ket{\CC_4}$ in
\begin{multline}
\WW_{\CC_4} :=3\cdot\openone^{\otimes 4} -\frac{1}{2}
\left(\sz^{(a)}\sz^{(b)}+\openone\right)
\left(\sz^{(b)}\sx^{(c)}\sx^{(d)}+\openone\right) \\
- \frac{1}{2}
\left(\sigma_x^{(a)}\sigma_x^{(b)}\sigma_z^{(c)}+\openone\right)
\left(\sigma_z^{(c)}\sigma_z^{(d)}+\openone\right), \label{CN2}
\end{multline}
with the theoretically optimal value of
$Tr(\WW_{\CC_4}\rho_{th})=-1$ \cite{Geza}. Thus, the correlations
in the two basis settings of Fig. 3 suffice to evaluate the
entanglement witness. Experimentally we find
$Tr(\WW_{\CC_4}\rho_{exp})=-0.299~\pm~0.050$ clearly proving the
genuine four-photon entanglement of the observed state.

In order to evaluate the quality of the state observed the
fidelity $F_{\CC_4} = \bra{\CC_4}\rho_{exp}\ket{\CC_4}$ is the
tool of choice. In general, full knowledge of the experimental
state, and therefore a complete state tomography would be
necessary to calculate the fidelity between two states. However,
again one can profit from the fact that the cluster state, as a
graph state, is completely describable by its stabilizers.
Therefore, the fidelity for the cluster state (as for any graph
states) equals the average expectation value of the stabilizer
operators. A measurement of the respective correlations is thus
sufficient to evaluate the state fidelity (Table \ref{tab:Stab}).
In our case, these are 16 correlations, instead of 81 for full
tomography, resulting in a value of $F_{\CC_4}=0.741 \pm 0.013$.


\begin{table}
\caption{\label{tab:Stab}Stabilizer Correlations}
\begin{ruledtabular}
\begin{tabular}{cc}
\begin{tabular}{r c@{$\otimes$}c@{$\otimes$}c@{$\otimes$}ccc}
&\multicolumn{4}{c}{operators} & \hspace{0.1\textwidth}
expectation value \hspace{0.1\textwidth}
\\[0.5ex]
(1)&$\phantom{-}\sz$   & $ \sz$   & $ \openone$ & $ \openone $ & $0.935 \pm 0.037$ \\
(2)&$\phantom{-}\sx$   & $ \sx$   & $ \sz$   & $ \openone$ & $0.713 \pm 0.044$ \\
(3)&$\phantom{-}\openone$ & $ \sz$   & $ \sx$   & $ \sx$   & $0.638 \pm 0.045$ \\
(4)&$\phantom{-}\openone$ & $ \openone$ & $ \sz$   & $ \sz$   & $0.931 \pm 0.036$ \\
(5)&$-         \sy$   & $ \sy$   & $ \sz$   & $ \openone$ & $0.679 \pm 0.043$ \\
(6)&$\phantom{-}\sz$   & $ \openone$ & $ \sx$   & $ \sx$   & $0.707 \pm 0.045$ \\
(7)&$\phantom{-}\sz$   & $ \sz$   & $ \sz$   & $ \sz$   & $0.931 \pm 0.064$ \\
(8)&$\phantom{-}\sx$   & $ \sy$   & $ \sy$   & $ \sx$   & $0.729 \pm 0.062$ \\
(9)&$\phantom{-}\sx$   & $ \sx$   & $ \openone$ & $ \sz$   & $0.673 \pm 0.044$ \\
(10)&$-          \openone$ & $ \sz$   & $ \sy$   & $ \sy$   & $0.626 \pm 0.067$ \\
(11)&$\phantom{-}\sy$   & $ \sx$   & $ \sy$   & $ \sx $  & $0.628 \pm 0.066$ \\
(12)&$-          \sy$   & $ \sy$   & $ \openone$ & $ \sz$   & $0.690 \pm 0.060$ \\
(13)&$-          \sz$   & $ \openone$   & $ \sy$   & $ \sy$ & $0.616 \pm 0.067$ \\
(14)&$\phantom{-}\sx$   & $ \sy$   & $ \sx$   & $ \sy $  & $0.681 \pm 0.066$ \\
(15)&$\phantom{-}\sy$   & $ \sx$   & $ \sx$   & $ \sy $  & $0.681 \pm 0.064$ \\
(16)&$\phantom{-}\openone$ & $ \openone$ & $ \openone$ & $ \openone$ & $1.00 \pm 0.017$ \\[0.7ex]
&\multicolumn{4}{c}{} & $F_{exp}=0.741\pm0.013$ \\
\end{tabular}
\end{tabular}
\end{ruledtabular}
\end{table}

The stabilizer correlations can be used as well for the
construction of a Bell inequality \cite{BI} with the following
Bell operator:
\begin{equation}
\SZ = \sz \openone \sx \sx + \sx \sy \sy \sx + \sx \sy \sx \sy -
\sz \openone \sy \sy \:.
\end{equation}
The maximal expectation value of $\SZ$ is obtained for the cluster
state giving the value $S=Tr( \SZ \rho_C)=4$, while the bound for
local hidden variable models is $S =2$. In our experiment we reach
$S=Tr( \SZ \rho_{exp})=2.73 \pm 0.12$ clearly violating the
classical bound.

Note, this Bell inequality is {\em not} violated by GHZ states,
which is a further indication of the cluster state showing a
different kind of entanglement compared to the GHZ state. Further
differences arise for the persistency of the entanglement, that is
the entanglement under projection or loss of particles
\cite{Persist_Briegel}.

A projective measurement onto $\sx$, which corresponds to a
projection onto the states $\ket{\pm 45^\circ}$ reduces both the
four-photon GHZ state and the cluster state to a three photon GHZ
state. Still, the resulting entanglement persistency is remarkably
different. Depending on the result of the projection measurement
(i.e. "$+45^\circ$" or "$-45^\circ$"), the four-photon GHZ state
reduces to $\ket{GHZ}_3^\pm=(\ket{HHH}\pm\ket{VVV})/\sqrt{2}$, the
incoherent sum of these two states exhibits no entanglement or
coherence whatsoever. Contrary, for the four-photon cluster state
we obtain, depending on the measurement result in mode $d$
\begin{eqnarray}\label{eq:oneout}\notag
\ket{\CC_3}_{abc}^{\pm} &=&
 \left(\ket{HH\pm}_{abc}+\ket{VV\mp}_{abc}\right)/\sqrt{2}\\ &=&
 \left(\ket{\Phi^+}_{ab}\ket{H}_c\pm\ket{\Phi^-}_{ab}\ket{V}_c\right)/\sqrt{2}.
\end{eqnarray}
According to the rules \cite{Persist_Briegel}, this is again a
cluster state, which can be further reduced, e.g. by projecting
the photon in mode $c$ onto $\ket{H}/\ket{V}$, to the two-photon
cluster states $\ket{\Phi^\pm}$. The incoherent sum of the states
of Eq. (\ref{eq:oneout}) gives the state $\rho_{a,b,c}$ which
results from the loss of photon $d$. This state still exhibits
two-partite entanglement.  To test these properties we apply the
entanglement witnesses for the respective states, which again
follow from the stabilizer formalism \cite{Geza} as
\begin{eqnarray}\notag
\WW_{\CC_{3,abc}^{\:\:\pm}}=\frac{3}{2}\cdot\openone^{\otimes
3}-\sx^{(a)}\sx^{(b)}\sz^{(c)}\\
-\frac{1}{2}\left(\sz^{(a)}\sz^{(b)}\openone^{(c)}\pm\sz^{(a)}\openone^{(b)}\sx^{(c)}
\pm\openone^{(a)} \sz^{(b)}\sx^{(c)}\right);
 \label{eq:GHZwitness}\\
\WW_{(\rho_{abc})}=\openone^{\otimes
3}-\sz^{(a)}\sz^{(b)}\openone^{(c)}-\sx^{(a)}\sx^{(b)}\sz^{(c)}\,.\label{eq:wit32}
\end{eqnarray}
In the experiment we observed the states with a fidelity of
$F_{\CC_{3,abc}^{\:\:+}}=0.756 \pm 0.028$ and
$F_{\CC_{3,abc}^{\:\:-}}=0.753 \pm 0.026$, yielding expectation
values of the entanglement witnesses of
$\langle\WW_{C_{3,abc}^{\:\:+}}\rangle=-0.362\pm 0.090$,
$\langle\WW_{C_{3,abc}^{\:\:-}}\rangle=-0.392\pm 0.082$, and
$\langle\WW_{\rho_{abc}}\rangle=-0.648\pm0.057$, respectively. We
obtain similar results for the witnesses when projecting and
tracing over other photons, indicating the high degree of
entanglement persistency of the cluster state.


\begin{figure}
\includegraphics[width=7cm]{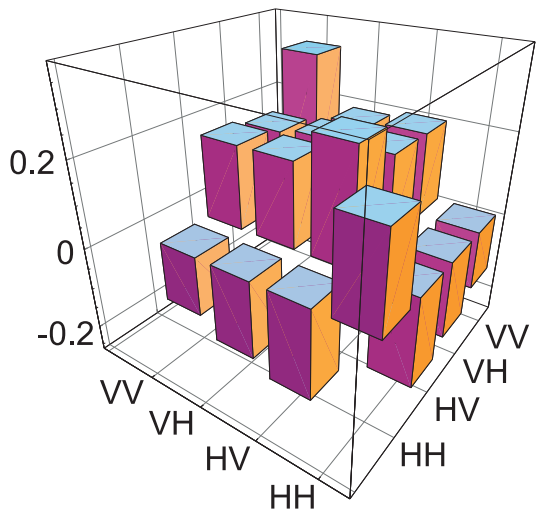}
\caption{State tomography after projection of photons in mode $b$
and $c$ onto $-45^\circ$. The state
$(\ket{H-}_{ad}-\ket{V+}_{ad})/\sqrt{2})$ is indeed obtained with
a fidelity of $0.809\pm0.027$.}
 \label{fig:2qubit}
\end{figure}

Finally, if we project two of the four photons onto suited bases,
we obtain two-photon entanglement for the remaining ones, which
reflects the maximum connectivity of the cluster state
\cite{Persist_Briegel}. In the scheme of the one-way quantum
computer, such a two-photon measurement corresponds to defining
the input values for a CNOT operation \cite{vienna}. For example,
if one projects/initializes photons in modes $b$ and $c$ to, say
both $\ket{-45^\circ}$, photons in modes $a$ and $d$ are due to
the CNOT plus Hadamard operation in the state
$(\ket{H-}_{ad}-\ket{V+}_{ad})/\sqrt{2})$. Fig. 4 shows the full
state tomography for this case. We experimentally obtain a
fidelity relative to the above state of $0.809\pm0.027$, resulting
in a logarithmic negativity \cite{negativity} for the observed
entanglement of $0.718\pm0.047$.

In summary we have presented a scheme for the preparation of a
four-qubit entangled cluster state in the polarization degree of
freedom of photons. This scheme is the first application of
 a new and simple way for the experimental realization of
a C-Phase gate based on linear optics.
The high stability and quality of the state creation enabled a
detailed experimental characterization of the entanglement
properties of a four-photon cluster state, including demonstration
of its genuine four-qubit entanglement and the study of
entanglement persistence under selective measurements and loss of
one or two qubits. We applied entanglement witnesses and
introduced a simplified fidelity analysis, which allows definite
characterization even without full state tomography. Such analysis
will become even more crucial in experiments with higher numbers
of qubits, where the effort for full state tomography increases
exponentially. The detailed study forms the basis to evaluate the
applicability of the experimental state for further quantum
information tasks, for example for multi-partite quantum
cryptography and one-way quantum computation.

This work was supported by the Deutsche Forschungsgemeinschaft and
the European Commission through the EU Project RamboQ
(IST-2001-38864).

\end{document}